\documentclass[12pt,a4paper,oneside]{article}
\usepackage{makeidx,ifthen,latexsym}
\usepackage{graphics,epsfig,rotating}
\newfont\bboard{msbm10}
\newfont\bboards{msbm7}
\newcommand{\kzeros}{$K^0_S$}

\newcommand{\yinel}{y}
\newcommand{\ystar}{y^*}
\newcommand{\av}[1]{\mbox{$ \left< #1 \right> $}}
\def\be{\begin{equation}}
\def\ee{\end{equation}}
\def\bea{\begin{eqnarray}}
\def\eea{\end{eqnarray}}
\def\ap#1#2#3   {{\em Ann. Phys. (NY)} {\bf#1} (#2) #3.}
\def\apj#1#2#3  {{\em Astrophys. J.} {\bf#1} (#2) #3.}
\def\apjl#1#2#3 {{\em Astrophys. J. Lett.} {\bf#1} (#2) #3.}
\def\app#1#2#3  {{\em Acta. Phys. Pol.} {\bf#1} (#2) #3.}
\def\ar#1#2#3   {{\em Ann. Rev. Nucl. Part. Sci.} {\bf#1} (#2) #3.}
\def\cpc#1#2#3  {{\em Comp. Phys. Comm.} {\bf#1} (#2) #3.}
\def\err#1#2#3  {{\it Erratum} {\bf#1} (#2) #3.}
\def\ib#1#2#3   {{\it ibid.} {\bf#1} (#2) #3.}
\def\jmp#1#2#3  {{\em J. Math. Phys.} {\bf#1} (#2) #3.}
\def\ijmp#1#2#3 {{\em Int. J. Mod. Phys.} {\bf#1} (#2) #3.}
\def\jetp#1#2#3 {{\em JETP Lett.} {\bf#1} (#2) #3.}
\def\jpg#1#2#3  {{\em J. Phys. G.} {\bf#1} (#2) #3.}
\def\mpl#1#2#3  {{\em Mod. Phys. Lett.} {\bf#1} (#2) #3.}
\def\nat#1#2#3  {{\em Nature (London)} {\bf#1} (#2) #3.}
\def\nc#1#2#3   {{\em Nuovo Cim.} {\bf#1} (#2) #3.}
\def\nim#1#2#3  {{\em Nucl. Instr. Meth.} {\bf#1} (#2) #3.}
\def\np#1#2#3   {{\em Nucl. Phys.} {\bf#1} (#2) #3.}
\def\pcps#1#2#3 {{\em Proc. Cam. Phil. Soc.} {\bf#1} (#2) #3.}
\def\pl#1#2#3   {{\em Phys. Lett.} {\bf#1} (#2) #3.}
\def\prep#1#2#3 {{\em Phys. Rep.} {\bf#1} (#2) #3.}
\def\prev#1#2#3 {{\em Phys. Rev.} {\bf#1} (#2) #3.}
\def\prl#1#2#3  {{\em Phys. Rev. Lett.} {\bf#1} (#2) #3.}
\def\prs#1#2#3  {{\em Proc. Roy. Soc.} {\bf#1} (#2) #3.}
\def\ptp#1#2#3  {{\em Prog. Th. Phys.} {\bf#1} (#2) #3.}
\def\ps#1#2#3   {{\em Physica Scripta} {\bf#1} (#2) #3.}
\def\rmp#1#2#3  {{\em Rev. Mod. Phys.} {\bf#1} (#2) #3.}
\def\rpp#1#2#3  {{\em Rep. Prog. Phys.} {\bf#1} (#2) #3.}
\def\sjnp#1#2#3 {{\em Sov. J. Nucl. Phys.} {\bf#1} (#2) #3.}
\def\spj#1#2#3  {{\em Sov. Phys. JEPT} {\bf#1} (#2) #3.}
\def\spu#1#2#3  {{\em Sov. Phys.-Usp.} {\bf#1} (#2) #3.}
\def\zp#1#2#3   {{\em Z. Phys.} {\bf#1} (#2) #3.}
\def\mylfigbox#1{%
\newbox\figbox
\setbox\figbox%
\hbox to 12cm{\hss\epsfysize=12cm\epsfbox{#1}\hss}
\hbox to \textwidth{\hfill \rotl\figbox\hfill}
}
\def\myltwoplots#1#2{%
\newbox\figboxone
\setbox\figboxone%
\hbox to 8cm{\hss\epsfysize=8cm\epsfbox{#1}\hss}
\newbox\figboxtwo
\setbox\figboxtwo%
\hbox to 8cm{\hss\epsfysize=8cm\epsfbox{#2}\hss}
\hbox to \textwidth{{\rotl\figboxone}\hfill{\rotl\figboxtwo}}
}
\def\myrfigbox#1{%
\newbox\figbox
\setbox\figbox%
\hbox to 8cm{\hss\epsfysize=8cm\epsfbox{#1}\hss}
\hbox to \textwidth{\hfill \rotr\figbox\hfill}
}
\def\twoplots#1#2{%
\hbox to \textwidth{\epsfysize=8cm\epsfbox{#1}\hfill\epsfysize=8cm\epsfbox{#2}}
}
\def\oneplot#1{%
\hbox to \textwidth{\hfill\epsfysize=12cm\epsfbox{#1}\hfill}
}
\def\emptypl{%
\hbox to \textwidth{\hfil\vbox to 8cm{\vfil\vskip8cm}}
}
\def\bigfigbox#1{%
\newbox\figbox
\setbox\figbox%
\hbox to 12cm{\hss\epsfysize=12cm\epsfbox{#1}\hss}
\hbox to \textwidth{\hfill \rotr\figbox\hfill}
}
\def\natplot#1{%
\hbox to \textwidth{\hfill \epsfbox{#1}\hfill}
}

\begin{document}
%-----------------------------\tableofcontents
%-------------------------------------------------------------------
\begin{titlepage}
\begin{flushleft}
{\tt DESY 97-095   \hfill  ISSN 0418-9833 } \\
{\tt May 1997       }\\
\end{flushleft}

\vspace*{4.cm}
\begin{center}
\begin{Large}
 
{\bf Photoproduction of \boldmath{$K^0$} and \boldmath{$\Lambda$}
 at HERA  and a  Comparison with Deep Inelastic Scattering}
 
\vspace{1.cm}
{H1 Collaboration}    \\
\end{Large}
\vspace*{1.5cm}

{\bf Abstract}
\end{center}
\begin{quotation}

\noindent
Inclusive $K^0$ and $\Lambda$ photoproduction
has been investigated at HERA with the H1 detector at an 
average photon-proton center of mass energy of 200~GeV in 
the transverse momentum range $0.5<p_t<5$~GeV.
The production rates as a function of $p_t$ and center of mass
rapidity are compared to those obtained in deep inelastic 
scattering at $\av{Q^2}=23$~GeV$^2$.
A similar comparison is made of the rapidity spectra of charged particles.
The rate of strangeness photoproduction is compared with $p\bar p$
measurements. The observations are also compared with next-to-leading
order QCD calculations and the predictions of a Monte Carlo model.

\end{quotation}
\end{titlepage}
%-------------------------------------------------------------------
%
\noindent
%   H1AUTS  Author list by names, no. of authors  392
%           status: 24/02/97   09.15.43
 C.~Adloff$^{35}$,                %WUPP-ST                  Adloff              
 S.~A\"\i d$^{13}$,               %HAM2-LEFT    8/96        Aid                 
 M.~Anderson$^{23}$,              %MANC-ST  10/95           Anderson            
 V.~Andreev$^{26}$,               %LPI -PD                  Andreev             
 B.~Andrieu$^{29}$,               %ECPL-PD                  Andrieu             
 V.~Arkadov$^{36}$,               %ZEUT-ST    10/96         Arkadov             
 C.~Arndt$^{11}$,                 %DESY-ST   1/96           Arndt               
 I.~Ayyaz$^{30}$,                 %PARI-ST       5/96       Ayyaz               
 A.~Babaev$^{25}$,                %ITEP-PD                  Babaev              
 J.~B\"ahr$^{36}$
% ...

\vspace*{0.5cm}
\newline
\noindent
%     H1 Institutes as appearing on publications
% ...
$ ^2$ III. Physikalisches Institut der RWTH, Aachen, Germany$^ a$ \\
% ...
$ ^{11}$ DESY, Hamburg, Germany$^a$ \\
% ...
$ ^{13}$ II. Institut f\"ur Experimentalphysik, Universit\"at Hamburg,
                                     Hamburg, Germany$^ a$  \\
% ...
$ ^{23}$ Physics Department, University of Manchester,
                                          Manchester, UK$^ b$\\
% ...
$ ^{25}$ Institute for Theoretical and Experimental Physics,
                                                 Moscow, Russia \\
 $ ^{26}$ Lebedev Physical Institute, Moscow, Russia$^ f$ \\
% ...
$ ^{29}$ LPNHE, Ecole Polytechnique, IN2P3-CNRS,
                             Palaiseau, France \\
 $ ^{30}$ LPNHE, Universit\'{e}s Paris VI and VII, IN2P3-CNRS,
                              Paris, France \\
% ...
$ ^{35}$ Fachbereich Physik, Bergische Universit\"at Gesamthochschule
               Wuppertal, Wuppertal, Germany$^ a$ \\
 $ ^{36}$ DESY, Institut f\"ur Hochenergiephysik,
                              Zeuthen, Germany$^ a$\\
% ...
$ ^a$ Supported by the Bundesministerium f\"ur Bildung, Wissenschaft,
        Forschung und Technologie, FRG,
        under contract numbers 6AC17P, 6AC47P, 6DO57I, 6HH17P, 6HH27I,
        6HD17I, 6HD27I, 6KI17P, 6MP17I, and 6WT87P \\
 $ ^b$ Supported by the UK Particle Physics and Astronomy Research
       Council, and formerly by the UK Science and Engineering Research
       Council \\
% ...
$ ^f$ Supported by the Deutsche Forschungsgemeinschaft \\
% ...

\newpage

%%%%%%%%%%%%%%%%%%%%%%%%%%%%%%%%%%%%
\section{Introduction}

The HERA collider allows the study of $ep$  interactions over a wide 
range of squared four-momentum transfer, $-Q^2$,
from photoproduction ($Q^2\simeq 0$) to very high photon
virtuality. This makes it possible to investigate the $Q^2$ dependence
of  particle production.

Photoproduction interactions are in many respects very similar to 
hadron-hadron interactions~\cite{bauer}. 
As $Q^2$ increases, the $ep$ interaction is considered 
to be deep inelastic scattering (DIS)
in which the exchanged virtual photon interacts directly with a parton 
in the proton.

The hadronic final state resulting from the fragmentation of partons
is of a non-perturba\-tive nature and is usually described using
phenomenological models such as the string fragmentation scheme
implemented in {\sc Jetset}~\cite{pythia}. In contrast 
to the study of unidentified charged particles~\cite{Eddie,Kuhlen}, 
strange particles allow tagging of a specific quark species. 
Strange quarks are mainly produced in the hadronization phase,
which is dominated by particles with low transverse momentum, $p_t$.
However, they can also originate from hard partonic radiation and the direct 
interaction of a photon with the proton, e.g. boson-gluon fusion,
thus contributing to the high-$p_t$ tail of strange particles.

In this paper, the transverse momentum and rapidity  
spectra of $K^0$ and $\Lambda$ particles in photoproduction
at an average center of mass energy $\av{W}=200$~GeV
are presented. These measurements are made in the photon fragmentation 
region, $1.3< \ystar < 2.8$, and 
the transverse momentum range $0.5 < p_t < 5$~GeV 
and are compared with results obtained in DIS 
at $\av{Q^2}=23$~GeV$^2$ and a similar average $W$. 
The comparison is extended to the rapidity spectrum of charged 
particles to see if the similarities observed for strange particles
are of a more general nature.
This allows the investigation of the $Q^2$ dependence of particle production
rates in the photon fragmentation region. In the central rapidity region
these rates are generally assumed to be independent of $Q^2$~\cite{bjkog}.
%which is generally assumed to hold for the central rapidity region, in the 
%photon fragmentation region.
The process dependence of particle production rates is studied by comparing 
the rates of strangeness production in $\gamma p$ and in $p \bar{p}$.
%The photoproduction results are also compared to {\sc Pythia}
%and to next-to-leading order (NLO) QCD calculations~\cite{NLO}.
Finally, results from next-to-leading order (NLO) QCD calculations~\cite{NLO} 
and from {\sc Pythia}~\cite{schuler,pythia} are compared with the measurements.

%%%%%%%%%%%%%%%%%%%%%%%%%%%%%%%%%%%%
\section{Models of Photoproduction}

The {\sc Pythia $5.7$} Monte Carlo program, as used here, generates
events according to the description given in~\cite{schuler,pythia}.
The structure of the photon is parameterized according
to~\cite{schuler,GRV-pion}.
Furthermore, the {\sc Pythia} option of generating  multiple
interactions within the same event is used~\cite{mia}.
Fragmentation is performed using 
the Lund string fragmentation scheme, as implemented in {\sc Jetset $7.4$}. 
Matrix elements are calculated to leading order,
and higher order terms are simulated by parton showers in the leading log
approximation. 
The production rate of strange particles in {\sc Jetset} is mainly
controlled by
the strangeness suppression factor, $\lambda_s$, that is, the probability
of producing a strange quark pair relative to a light quark pair. 
By default $\lambda_s$ is set to $0.3$. 
However, recent deep inelastic muon nucleon scattering data from 
E665~\cite{E665}, $e^+e^-$ data from DELPHI~\cite{DELPHI} and $ep$ data from 
H1~\cite{h1-DISk0} and ZEUS~\cite{zeus-DISk0} are better described by
{\sc Jetset} if  $\lambda_s\simeq0.2$~\footnote{The following
{\sc Jetset} hadronization parameters have been changed with respect
to the default settings (default/DELPHI/E665):
Parj(2)=$\lambda_s$=(0.3/0.23/0.2), Parj(11)=(0.5/0.365/0.5), 
Parj(12)=(0.6/0.41/0.6).
Parj(2) is the strangeness suppression factor, Parj(11) is the
probability that a meson containing $u$ or $d$ quarks has spin 1,
Parj(12) the probability that a meson containing $s$ quarks has spin 1.}.

The authors of  \cite{NLO} calculate the inclusive 
single-particle photoproduction cross sections as follows.
The hard subprocess is calculated to
next-to-leading order in QCD, as is the evolution of the fragmentation
and parton distribution functions. The
fragmentation functions are derived from $e^+e^-$ collider data.
The parton distributions for the photon and the proton
constitute an important input to the cross-section calculations.
The numerical calculations  presented here  are based on the
GRV parameterizations given in \cite{GRV-photon/proton}.

%%%%%%%%%%%%%%%%%%%%%%%%%%%%%%%%%%%%

\section{Experimental Procedure}

\subsection{Detector}

The results presented in this paper are based on data taken in 1994
with the H1 detector. During this running period 820~GeV protons and
27.5~GeV positrons were brought into collision in HERA.

A full description of the H1 detector can be found in \cite{h1det}.
In this paper  only  the  components  of relevance to the
measurements presented here are mentioned. These are
the positron tagger, the central tracker, the backward proportional chamber 
(BPC), the backward electromagnetic calorimeter (BEMC) and the liquid argon 
(LAr) calorimeter.
The direction of the $z$-axis is chosen to be along the proton beam direction.
The polar angle $\theta$ is defined with respect to the $z$-axis and the 
pseudorapidity is given by $\eta = -\ln(\tan\theta/2)$.  
 
The positron tagger, located
$33\,\hbox{m}$ from the interaction point in the positron direction,
measures the energy of the scattered positron and, in conjunction with 
a photon detector,  the luminosity by exploiting the 
Bethe-Heitler process \cite{h1lumi}.

The central track detector consists of an inner ($20 <R<45\,\hbox{cm}$) 
and an outer ($53 <R<85\,\hbox{cm}$) concentric central jet chamber (CJC),
multi-wire proportional chambers for triggering purposes 
and two additional drift chambers which measure accurately the
$z$-coordinate.
The pseudorapidity range covered by the central track detector is $|\eta|<1.5$.

The LAr calorimeter, which surrounds the tracking system, consists of
an electromagnetic and a hadronic section. Here it is only used
to identify and reject events containing a large rapidity gap in the
forward direction, in the range $2.03 < \eta < 3.26$.

In the backward direction the BPC is used to determine the polar angle
and the BEMC the energy of the scattered positron in DIS events.
The pseudorapidity range covered by these detectors is
$-3.35 < \eta < -1.51$.

The calorimeters are surrounded by a superconducting coil providing a uniform
magnetic field of 1.15~T in the region occupied by the central tracker.

\subsection{Event Selection}

Photoproduction events are selected by requiring the presence of a scattered
positron in the small angle positron tagger.
For these events $Q^2 < 10^{-2}$~GeV$~^2$. 
The inelasticity of the scattering is given by
$\yinel  = 1- \frac{E^\prime_e}{E_e} \cdot \sin^2\theta/2$.
Here, $E_e$ is the energy of the incident positron, $E^\prime_e$ and $\theta$
the energy and the polar angle of the scattered positron.  
In photoproduction, to a very good approximation, the inelasticity is given by
$ \yinel  = 1- \frac{E^\prime_e}{E_e}$.
Photoproduction events are required to lie in
the range $0.3 < \yinel < 0.7$, in order to confine the scattered positron
to a region in which the acceptance of the positron tagger is well understood.
The trigger for these events requires a signal in the positron tagger and
the presence of at least one negative track in the CJC.
More information on this track-based trigger can be found in~\cite{t83}.

The trigger and the event selection for the DIS sample
are as described in~\cite{h1-DISk0}
except that the $y$ range is restricted to ensure that the mean
hadronic center of mass energies of the DIS and photoproduction samples
are similar. The scattered positron is detected in the BEMC
and is required to have an energy greater than 12~GeV and 
a polar angle between 150$^\circ$ and 173$^\circ$. 
The kinematic variables are determined using the
measured energy and angle of the scattered positron.
The DIS sample is restricted to and corrected in the kinematic region
given by $10\leq Q^2 \leq 70$~GeV$^2$, $10^{-4}\leq x \leq 10^{-2}$
and $0.3\leq \yinel \leq 0.6$, where $x$ is the Bjorken scaling variable.

For both the $\gamma p$ and DIS samples,  the corresponding 
integrated luminosity  is 1.3~pb$^{-1}$.

In addition to the measurements of strangeness production in $\gamma p$ 
and DIS, results on charged-particle production in these two processes are 
presented.
For this purpose, a charged-particle photoproduction sample is selected
consisting of events triggered by the coincident
detection of a positron in the small angle positron tagger and 
at least one track pointing to the interaction vertex with $p_t>200$~MeV 
measured in the cylindrical multi-wire proportional chambers.
The charged-particle DIS sample consists of events which are triggered by 
a cluster in the BEMC with an energy deposit of more than 4~GeV.
For both the DIS and $\gamma p$ charged-particle samples,
the inelasticity $\yinel$ of the interaction is restricted  
to $0.3 < \yinel < 0.5$, resulting in an average
$W$ of 187~GeV. The DIS sample is restricted to the range
$8 < Q^2 < 30$~GeV$^2$ 
giving an average $\av{Q^2}=15$~GeV$^2$.
The charged-particle data samples studied correspond to an integrated 
luminosity of 1.5~pb$^{-1}$. 

Common to all four event samples is the requirement of a reconstructed primary 
vertex within 30~cm of the nominal vertex position.

The background from photoproduction in the two DIS data samples is estimated
to be less than 3\%. 

For a comparison of particle production in different processes it is important
to take into account the different contributions of low-mass diffractive 
events.
While in the case of DIS this fraction is only 10\%, it is around 30\% in 
photoproduction~\cite{h1-diffr-gp,h1-diffr-DIS}.
These events are characterized by the absence of energy deposited in the 
forward direction. To allow a direct comparison between DIS and
$\gamma p$ these large rapidity gap events are removed from the samples by
requiring the presence of at least 500~MeV of energy in the forward region
of the liquid argon calorimeter, $2.03 < \eta < 3.26$.
The remaining samples are termed non-diffractive in the following.
 
%
%-------------------------------------------------
\subsection{ \boldmath{$K^0_S$} and \boldmath{$\Lambda$}
 Identification in Photoproduction}
%-------------------------------------------------
%
%
$K^0_S$ mesons and $\Lambda$ baryons are identified through their decay 
channels 
\[
 K^0_S \rightarrow \pi^+\pi^-    
\]
\[
 \Lambda ({\bar\Lambda})\rightarrow p \pi^{-} ({\bar p} \pi^{+})
\]
where the pion, the proton and the antiproton 
tracks are reconstructed in the CJC.
\begin{figure}[hhhhh]
\begin{center}
\epsfig{file=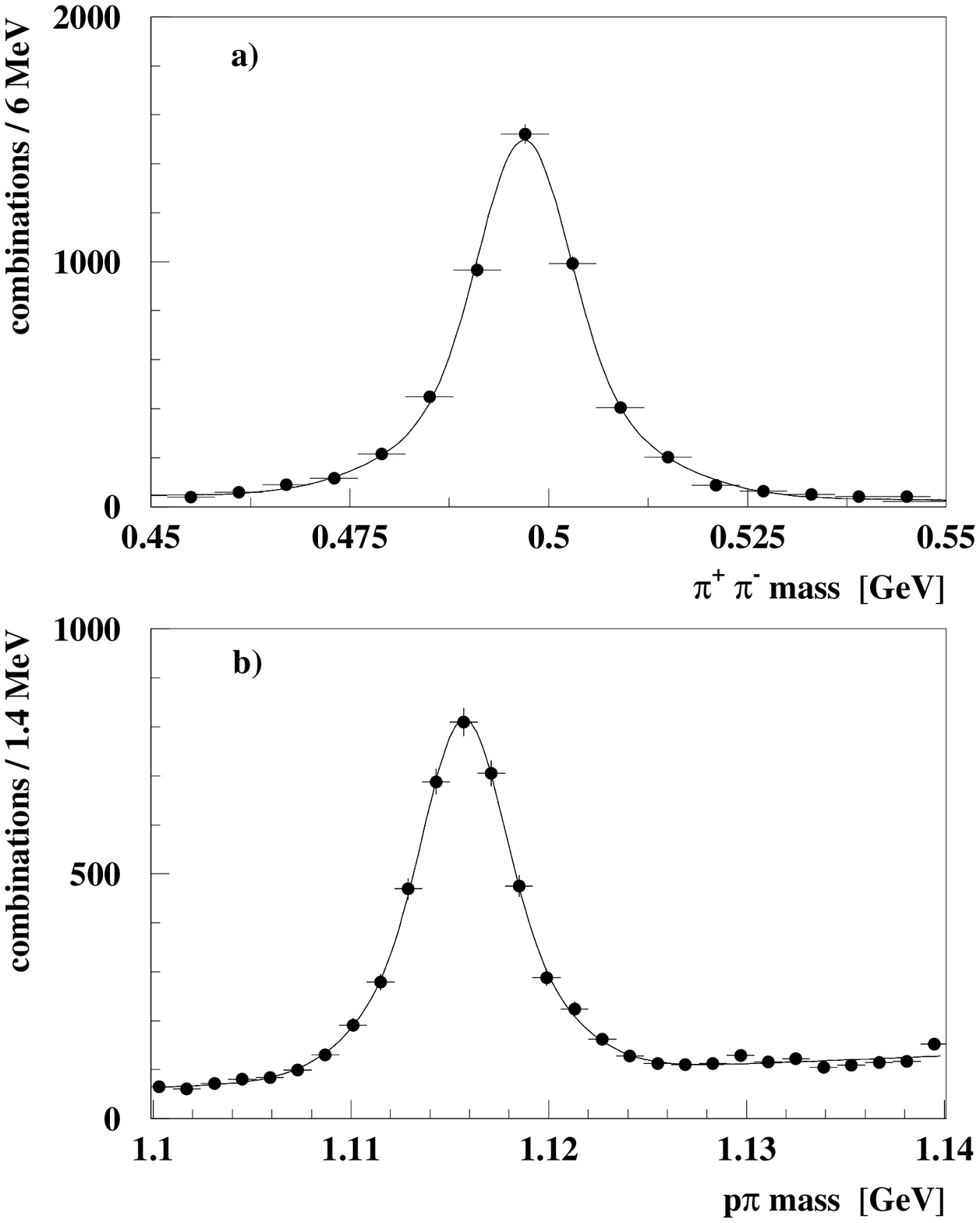,
              height=18cm}
\caption{{  
Invariant mass distributions for
$K^0_S$~(a) and $\Lambda$~(b) candidates in photoproduction, after applying
all the cuts described in the text. The $K^0_S$ candidates are restricted to 
the kinematic range $|\eta|<1.3$ and $0.5<p_t<5$~GeV; the $\Lambda$ candidates
are restricted to the kinematic range $|\eta|<1.3$ and $0.6<p_t<5$~GeV.
The curves represent the fits as described in the text.}}

\label{f:ksignal}
\end{center}
\end{figure}
The tracks of the decay products are required to satisfy $|\eta| < 1.5$ and
$p_t > 180\,\hbox{MeV}$, where $p_t$ is defined with respect to  
the beam axis. This restricts the acceptance to a region
where the reconstruction efficiency is high.
In addition, it is  required that the radial positions $r_{beg}$ and
$r_{end}$ of the first and last  measured point on the tracks fulfill 
$r_{beg}<30\,\hbox{cm}$ and $r_{end}>37.5\,\hbox{cm}$ and that  their
radial length be at least $10\,\hbox{cm}$.
Candidates for neutral particles decaying into two charged particles,
hereafter called $V^0$ candidates, are searched for by performing a
constrained fit to each pair of oppositely charged tracks. The fit 
demands that the tracks meet at a common secondary vertex and that the two
particle momenta of the decay products transverse to the path of flight of
the $V^0$ candidate be opposite.
The $V^0$ candidate invariant masses are calculated assuming that the
decay products are two pions for $K^0_S$ candidates and a proton and a pion for
$\Lambda$ candidates.

To reduce the combinatorial background and ensure good reconstruction of
the decay, the radial distance $d_r$ from the primary
to the secondary vertex is restricted to $2 < d_r < 18\,\hbox{cm}$ for the
$K^0_S$ and $3 < d_r < 18\,\hbox{cm}$ for the $\Lambda $ candidates. 
Note that $\Lambda $ decays  
with $d_r<3$~cm have a lower reconstruction probability due to the large
asymmetry in the $p_t$ of the decay particles.

The contamination from $\Lambda$ decays in the $K^0_S$ sample
is reduced to a negligible level by demanding that the
transverse momentum 
of the decay particles with respect to the momentum of the $V^0$ 
candidate be greater than $120\,\hbox{MeV}$.
Likewise, the $K^0_S$ contamination in the $\Lambda$ sample is 
removed by excluding those $\Lambda$ candidates for which the
$\pi^+\pi^-$  invariant mass, $m_{\pi\pi}$, falls in the window
$0.48 < m_{\pi\pi} < 0.52$~GeV.

The applied cuts ensure that kinematic regions in which the reconstruction
efficiencies for $K^0_S$ and $\Lambda$ start to degrade are avoided.
The large number of $K^0_S$ candidates makes it possible
to exclude in addition  regions in azimuth, $\varphi$, of the CJC which have
known imperfections (and where systematic effects are larger)
and to reject $K^0_S$ topologies which are dominated by combinatorial 
background \cite{karendiss}. 

In order to ensure an optimal acceptance of the CJC for the $V^0$'s, 
their pseudorapidity
has been restricted to $|\eta|<1.3$ and their transverse momentum
to $0.5<p_t<5$~GeV for \kzeros\ and to $0.6<p_t<5$~GeV
for $\Lambda$.

The final $K^0_S$ and $\Lambda$ signals, after applying all the aforementioned
cuts, are shown in Fig.~\ref{f:ksignal}. Approximately $7\,700$ $K^0_S$
candidates and $3\,600$ $\Lambda$ candidates remain.
In the fits the signals are described by a superposition of two gaussians
on a linear background, accounting for
varying invariant mass resolutions for different decay topologies. 
On average the RMS width  of the $K^0_S$ signal
is 9.2~MeV, the RMS width of the $\Lambda$ signal is $2.9$~MeV.
The fitted peak positions
are in agreement with the PDG  $K^0$ and $\Lambda$ masses~\cite{PDG} within the
systematic errors resulting from performing the above fits using various
functional forms and from
remaining uncertainties in the calibration of the tracking system.

\subsection{Corrections and Systematic Errors}

In order to obtain the inclusive $K^0$ and $\Lambda$ photoproduction
cross sections it is necessary to correct the observed numbers
of $K^0_S$ mesons and  $\Lambda$ baryons
for the branching ratios into the observed decay channels, and
for the acceptances and efficiencies of the various detector components.

The dependence of the acceptance of the positron tagger on
the inelasticity, $y$, is known from a measurement of
the tagger acceptance for the Bethe-Heitler process $ep\rightarrow ep\gamma$
and Monte Carlo studies of the luminosity system
and the HERA beam optics. More details can be found in~\cite{totalgp}. 

The efficiency of the track trigger is studied by using
a reference event sample obtained with an
independent trigger which does not make any requirement
on the hadronic final state. On average it is found to
be 85\% and only weakly dependent on the transverse
momentum of the $V^0$. 
The systematic uncertainty associated with the trigger efficiency
is less than $6\%$. This has been cross-checked by using a sample of 
simulated and reconstructed events. 
  
A visual check of the efficiency for reconstructing tracks satisfying
the requirements described in the previous section reveals that it is
consistent with the estimate of 98\% obtained in \cite{erdmann}.
The systematic uncertainty assigned to this correction factor is 2\%.

The main correction results
from the limited geometric acceptance of the CJC for $V^0$ decays
and the cuts applied to select $V^0$ candidates explained in the 
previous section. The acceptance can be parameterized as a function
of the $p_t$ and $\eta$ of the $V^0$ and ranges between
3\% and 17\% for $K_S^0$ and between 4\% and 25\% for $\Lambda$ decays.
The cuts which have a large effect on the acceptance are the lower cuts 
on $d_r$ and $p_t$ of the decay tracks and, in the case of the $K^0_S$, the 
$\varphi$ cut. The bin-dependent systematic uncertainty on the acceptance 
calculation is estimated by varying all cuts independently and re-evaluating
the cross section.
In a given bin the maximum variation is taken as a measure of the systematic
error resulting from the acceptance correction.
The bin-dependent systematic errors are listed in Tab. 1 and 2 and  
range between 2\% and 10\%.

Within the region defined by the above cuts the efficiency for reconstructing
$V^0$'s is found to be 
100\% using a full detector simulation of photoproduction events.
Since the $K^0$ analysis is restricted to the fully efficient regions
of the CJC, which are well simulated, no additional correction is
applied. For the reconstruction of $\Lambda$ decays a detection efficiency of 
97\% is found.
This leads to a systematic error of $^{+2}_{-0}\%$ for $K^0_S$
and $^{+5}_{-3}\%$ for $\Lambda$. 

The background resulting from
non-{\em ep} interactions (proton wall and proton gas interactions) has
been estimated from the  proton pilot-bunch events, that is,
events where the proton bunch does not have a colliding positron bunch partner.
This background has been found to be negligible, but a
systematic uncertainty of $^{+0}_{-2}\%$ is assigned to its contribution
to the cross section.

The systematic uncertainty of the luminosity determination is 
1.5\%~\cite{lumi}.

The overall bin-independent systematic error is calculated by adding the
various contributions in quadrature and results in errors of $\pm 8\%$ for 
the $K^0$ cross section and $^{+9}_{-8}\%$ for the $\Lambda$ cross section.

For the charged-particle spectra, the dominant contributions to the
systematic error, both in
photoproduction and DIS, originate from the extrapolation to $p_t=0$ (5\%).
The extrapolation was performed using a sample of fully simulated Monte Carlo
events. The resulting error was estimated by varying the minimum $p_t$ of the
particles both in Monte Carlo and data.
In the case of the total charged-particle photoproduction sample an additional
uncertainty in the losses of events with low charged-particle multiplicity 
(6\%) is taken into account.

%%%%%%%%%%%%%%%%%%%%%%%%%%%%%%%%%%%%
%---------------------------------
\section{Experimental Results}
%---------------------------------
%
Comparisons between the photoproduction data and the above-mentioned
models and calculations are performed in the laboratory frame. 
Comparisons of strangeness production in photoproduction with that in
DIS and $p\bar{p}$ interactions are made in the
photon-proton and the proton-antiproton center of mass systems, 
using the non-diffractive data sample.
The transverse momentum, $p_t$, and the rapidity, $\ystar$, are defined with
respect to the direction of the exchanged boson for DIS and
with respect to the direction of the positron beam  for $\gamma p$.
In the case of $p\bar p$,  $p_t$ and $| \ystar |$ are defined with respect to 
the  beam axis. 
 
In Tab.~\ref{t:dpt2dystar} and \ref{t:dystar} 
the total $\gamma p$ cross sections for $K^0$, $\Lambda$ and $\bar{\Lambda}$ 
production are presented as a function of $p_t$ and rapidity $\ystar$.
The $\ystar$ range has been restricted to $1.3< \ystar< 2.8$ to make the
measurement insensitive to the $\theta$ acceptance of the CJC.
The $p_t$ spectra fall steeply whereas there is almost no dependence
on $\ystar$ in the measured rapidity interval.

Here and in what follows, all $K^0_S$ results are multiplied by
2 and therefore correspond to $K^0$ and $\bar{K^0}$ production; no distinction
is made between  $K^0$ and $\bar{K^0}$ and the combination of both is
denoted by $K^0$.
Furthermore, it can be seen that the $\Lambda$ cross section is, within errors,
equal to the $\bar \Lambda$ cross section. Therefore they are combined and 
subsequent references to $\Lambda$ always mean $\Lambda + \bar \Lambda$.

For photoproduction the $e p\rightarrow V^0 X$ cross section $\sigma_{ep}$
is related to the $\gamma p\rightarrow V^0 X$ cross section
$\sigma_{\gamma p}(W)$ by the Weizs\"acker-Williams formula~\cite{flux}
$$\sigma_{ep} = \sigma_{\gamma p}(W) \cdot F$$ 
where $F=0.0136$ is the flux factor integrated over $0.3<\yinel <0.7$ and 
$Q^2<10^{-2}$~GeV$^2$.
\begin{table}[hhhh] 
\begin{center}
\vbox{
\begin{tabular}{c|cc|rrr|} 

\cline{2-6}  
 & \multicolumn{2}{|c|}{}& {\Large{ $ \frac{1}{\Delta \ystar}
\frac{d\sigma}{dp_t^2}$ }} &
 stat. & syst. \\
 & \multicolumn{2}{|c|}{$p_t$ range}&  &  &  \\
 & \multicolumn{2}{|c|}{ [GeV]}& \multicolumn{3}{c|}
{  [nb/GeV$^2$]} \\
%\cline{2-6} 
%\multicolumn{6}{c}{}\\
\cline{2-6} 
a) {\bf $K^0$:}\hspace{5mm}
    & 0.5& 0.6 &$25500  $&$1400  $ &$500  $\\
    & 0.6& 0.7 &$16700  $&$ 900  $ &$300  $\\
    & 0.7& 0.8 &$11300  $&$ 600  $ &$200  $\\
    & 0.8& 0.9 &$ 7800  $&$ 400  $ &$300  $\\
    & 0.9& 1.0 &$ 4900  $&$ 300  $ &$200  $\\
    & 1.0& 1.2 &$ 3200  $&$ 100  $ &$200  $\\
    & 1.2& 1.4 &$ 1600  $&$ 100  $ &$ 50  $\\
    & 1.4& 1.8 &$  570  $&$  30  $ &$ 30  $\\
    & 1.8& 2.2 &$  190  $&$  20  $ &$ 10  $\\
    & 2.2& 3.0 &$   38  $&$   5  $ &$  3  $\\
    & 3.0& 5.0 &$    3.6$&$   0.9$ &$  0.4$\\
\cline{2-6}  
\multicolumn{6}{c}{}\\
\cline{2-6}  
b) {\bf $\Lambda$:}\hspace{5mm}
%    & 0.6 & 0.9 &$1786.2$&$115.8$ &$105.7$\\
%  scaled by 1.0421 due to effect of eta cut  15.4.97
    & 0.6 & 0.9 &$1900$&$100$ &$100$\\
    & 0.9 & 1.2 &$ 630$&$ 30$ &$ 40$\\
    & 1.2 & 1.7 &$ 190$&$ 10$ &$ 20$\\
    & 1.7 & 3.0 &$  17$&$  2$ &$  1$\\
\cline{2-6}
\multicolumn{6}{c}{}\\
\cline{2-6}  
c) {\bf $\overline{\Lambda}$:}\hspace{5mm}
%    & 0.6 & 0.9 &$1671.3$&$111.5$ &$98.9$\\
%  scaled by 1.0363 due to effect of eta cut  15.4.97
    & 0.6 & 0.9 &$1700$&$100$ &$100$\\
    & 0.9 & 1.2 &$ 600$&$ 30$ &$ 30$\\
    & 1.2 & 1.7 &$ 200$&$ 10$ &$ 20$\\
    & 1.7 & 3.0 &$  15$&$  2$ &$  1$\\
\cline{2-6}
\end{tabular}

\vfill}
\caption{ Inclusive $K^0$ (a) and $\Lambda/\overline{\Lambda}$ (b,c)
 total $\gamma p$ production cross sections 
 $1/\Delta \ystar \cdot d\sigma/dp_t^2$ for
 $1.3<\ystar<2.8$. The systematic
 errors given in the table do not include an overall systematic uncertainty of
 $\pm 8\%$ for $K^0$ and $^{+9}_{-8}\%$ for $\Lambda$ as explained in section 
 3.4. }
\label{t:dpt2dystar} 
\end{center} 
\end{table}
\begin{table}[hhhh] 
\begin{center}

\begin{tabular}{c|cc|rrr|} 

\cline{2-6}  
 & \multicolumn{2}{|c|}{}& {\Large{ $ \frac{d\sigma}{d \ystar }$ }} &
 stat. & syst. \\
 & \multicolumn{2}{|c|}{$\ystar$ range} &   & [nb] & \\
\cline{2-6} 
a) {\bf $K^0$:}\hspace{5mm}
    & $1.30$&$1.55$&$13500$&$600$ &$700$\\
    & $1.55$&$1.80$&$12700$&$600$ &$400$\\
    & $1.80$&$2.05$&$13200$&$600$ &$300$\\
    & $2.05$&$2.30$&$11800$&$600$ &$500$\\
    & $2.30$&$2.55$&$12600$&$600$ &$400$\\
    & $2.55$&$2.80$&$12200$&$600$ &$400$\\
\cline{2-6}  
\multicolumn{6}{c}{}\\
\cline{2-6}  
b) {\bf $\Lambda$:}\hspace{5mm}
%    & $1.300$&$1.675$&$1746$&$111$ &$131$\\
% scaled by 1.0827 due to effect of eta cut  15.4.97
    & $1.300$&$1.675$&$1900$&$100$ &$100$\\
    & $1.675$&$2.050$&$1600$&$100$ &$100$\\
    & $2.050$&$2.425$&$1600$&$100$ &$300$\\
%    & $2.425$&$2.800$&$1500$&$110$ &$70$\\
% scaled by 1.0195 due to effect of eta cut  15.4.97
    & $2.425$&$2.800$&$1500$&$100$ &$100$\\
\cline{2-6}
\multicolumn{6}{c}{}\\
\cline{2-6}  
c) {\bf $\overline{\Lambda}$:}\hspace{5mm}
%    & $1.300$&$1.675$&$1813$&$135$ &$150$\\
% scaled by 1.0701 due to effect of eta cut  15.4.97
    & $1.300$&$1.675$&$1900$&$100$ &$100$\\
    & $1.675$&$2.050$&$1600$&$100$ &$100$\\
    & $2.050$&$2.425$&$1300$&$100$ &$200$\\
%    & $2.425$&$2.800$&$1344$&$103$ &$63$\\
% scaled by 1.0162 due to effect of eta cut  15.4.97
    & $2.425$&$2.800$&$1400$&$100$ &$100$\\
\cline{2-6}
\end{tabular}

\caption{   Inclusive $K^0$ (a) and $\Lambda/\overline{\Lambda}$ (b,c)
 total $\gamma p$ production cross sections 
 $d\sigma/d\ystar$ for $0.5<p_t<5$~GeV in the case of $K^0$'s
 and $0.6<p_t<5$~GeV in the case of $\Lambda$'s. The systematic
 errors given in the table do not include an overall systematic uncertainty of
 $\pm 8\%$ for $K^0$ and $^{+9}_{-8}\%$ for $\Lambda$ 
 as explained in section~3.4. }
\label{t:dystar} 
\end{center} 
\end{table}

\subsection{Comparison with Models and Calculations}

In this section the predictions of
{\sc Pythia} in combination with {\sc Jetset},
and the results of next-to-leading order calculations, are
confronted with the measurements reported in this paper.

\begin{figure}
\begin{center}
\epsfig{file=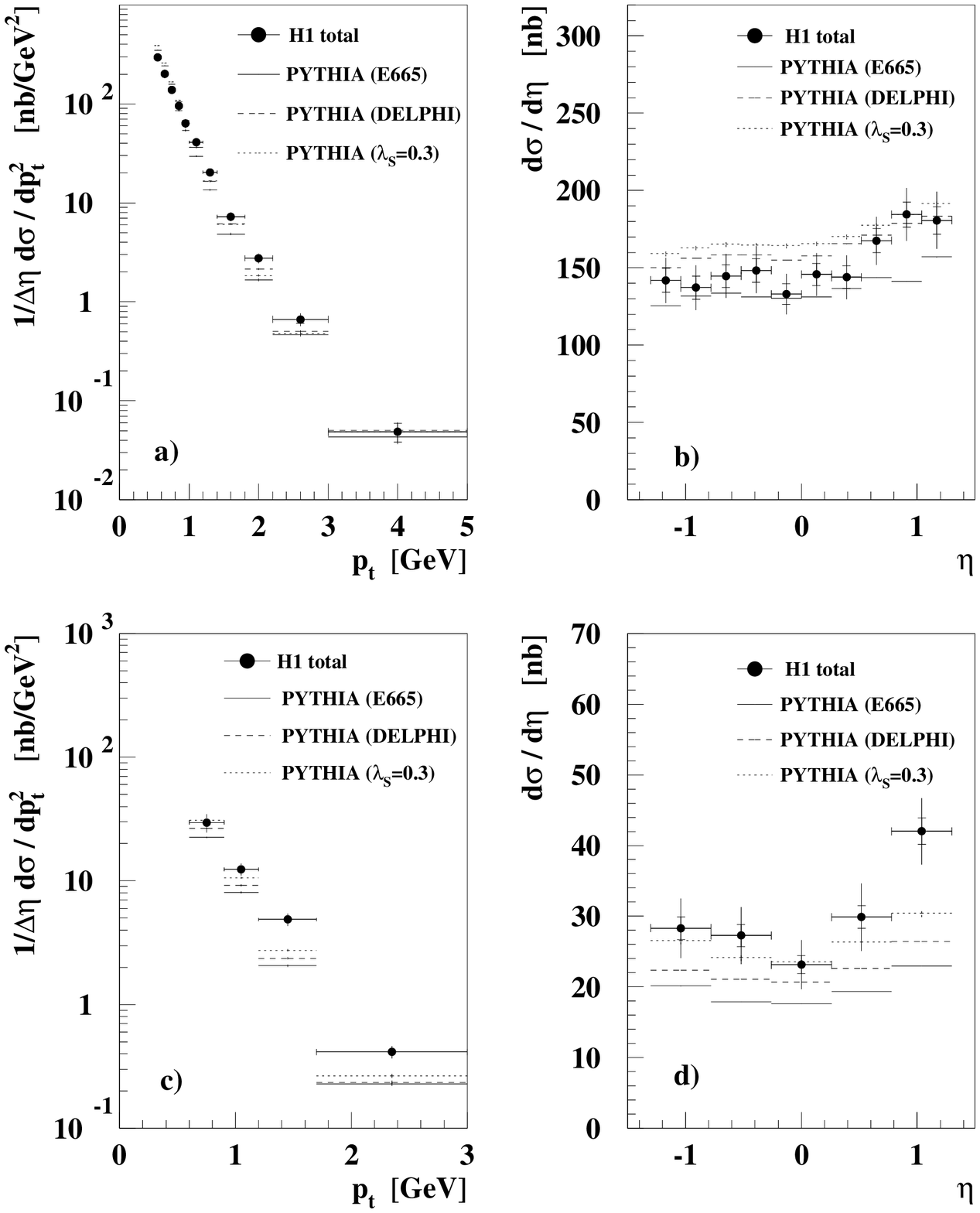,
              width=14cm}
\caption{  Measured $K^0$ (a,b)  and $\Lambda$ (c,d) total $ep$ cross sections
in photoproduction as a function of $p_t$  and pseudorapidity $\eta$ 
compared to {\sc Pythia} in the kinematic range $|\eta|<1.3$ and 
$0.5<p_t<5$~GeV.
The inner vertical error bars indicate the statistical error only and the outer
error bars the statistical and total systematic error added in quadrature.
The labels E665 and DELPHI correspond to {\sc Jetset} parameters
favored by these experiments; the label $\lambda_s =0.3$ denotes standard 
{\sc Jetset} settings. }
\label{f:data-mc}
\end{center}
\end{figure}

\begin{figure}
\begin{center}
\epsfig{file=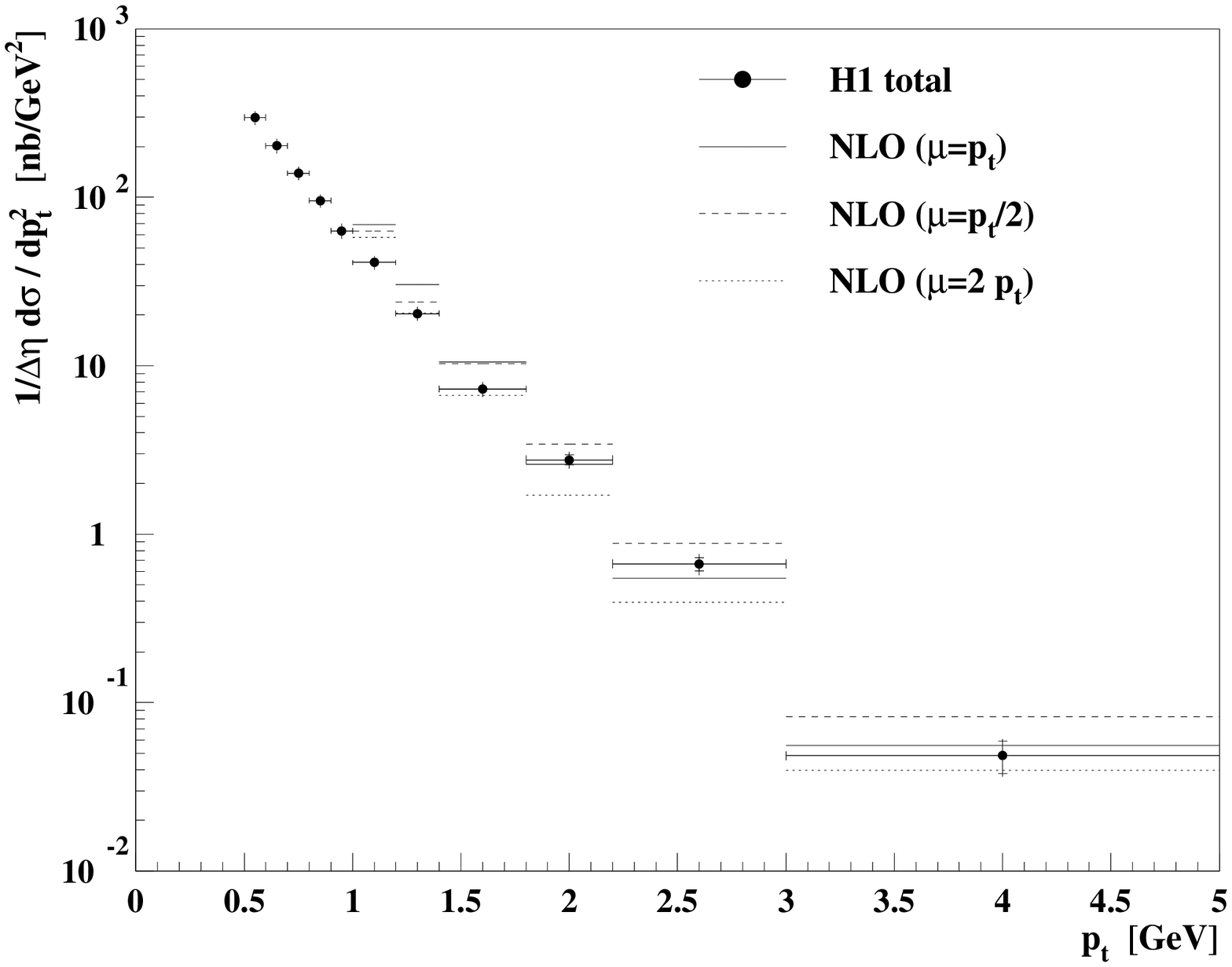,
              width=16cm}
\caption{ Measured $K^0$ total $ep$ cross section in photoproduction 
 as a function of $p_t$ compared to  
 NLO calculations in the kinematic range
 $|\eta|<1.3$ and $0.5<p_t<5$~GeV.
 The inner vertical error bars indicate the statistical error only and the 
 outer error bars the statistical and total systematic error added in 
 quadrature. The NLO curves correspond
 to different values of the renormalization and factorization scales $\mu$, 
 namely $p_t$, $2p_t$ and $p_t/2$, where $p_t$ is the transverse momentum 
 of the $K^0$.  }
\label{f:data-nlo}
\end{center}
\end{figure}

Figure~\ref{f:data-mc} shows a comparison of the measured
$K^0$ and $\Lambda$ cross sections in $p_t$ and $\eta$ with the predictions
of {\sc Pythia} using standard parameter settings ($ \lambda_s=0.3$) as well
as parameter sets obtained from fits to DELPHI and E665 data, as described 
in section~2.
In the case of $K^0$ mesons (Fig.~\ref{f:data-mc}~a,b) the measured cross 
sections are in reasonable agreement with the Monte Carlo predictions using 
the DELPHI or E665 settings whereas the predictions with the standard settings
seem to overestimate the cross section.
However, the measured $\Lambda$ (Fig.~\ref{f:data-mc}~c,d) cross section lies
significantly above the Monte Carlo prediction for all three scenarios.
The UA5 collaboration has reported a similar discrepancy between
$\Lambda$ production in $p\bar p$ interactions at 200~GeV and {\sc Pythia}
predictions~\cite{ua5_lambda}.
No attempt is made here to tune the {\sc Jetset} parameters
controlling diquark production.

Recent next-to-leading order QCD calculations of inclusive particle spectra 
in $ep$ scattering use fragmentation functions which have been fitted to 
$e^+e^-$ data~\cite{NLO}.
These NLO calculations involve four scales:
the factorization scales connected with the parton densities of the photon
and the proton, the factorization scale of the fragmentation functions and
the renormalization scale of the QCD coupling constant.
The calculations for $K^0$ cross sections assume that all four scales are
equal; they are taken to be equal to $p_t/2$, $p_t$ or $2p_t$ (where
$p_t$ is the transverse momentum of the $K^0$).
This arbitrariness in the choice of the scales introduces a change in the 
predicted cross sections and is a measure of the contribution of
higher-order terms.
To reduce the scale influence, the calculation for
the photoproduction of neutral kaons was restricted to transverse $K^0$
momenta  $p_t > 1$~GeV, although the theoretical predictions
should be more reliable for transverse momenta above 3~GeV~\cite{NLO}.
A comparison of data with this calculation is shown in Fig.~\ref{f:data-nlo}.
In the high $p_t$ region the agreement is satisfactory. Below 2~GeV the
calculation deviates significantly from the data. 

\subsection{Comparison to Deep Inelastic Scattering and \boldmath{$p\bar{p}$}}

\begin{figure}[hhhh]
\begin{center}
\epsfig{figure=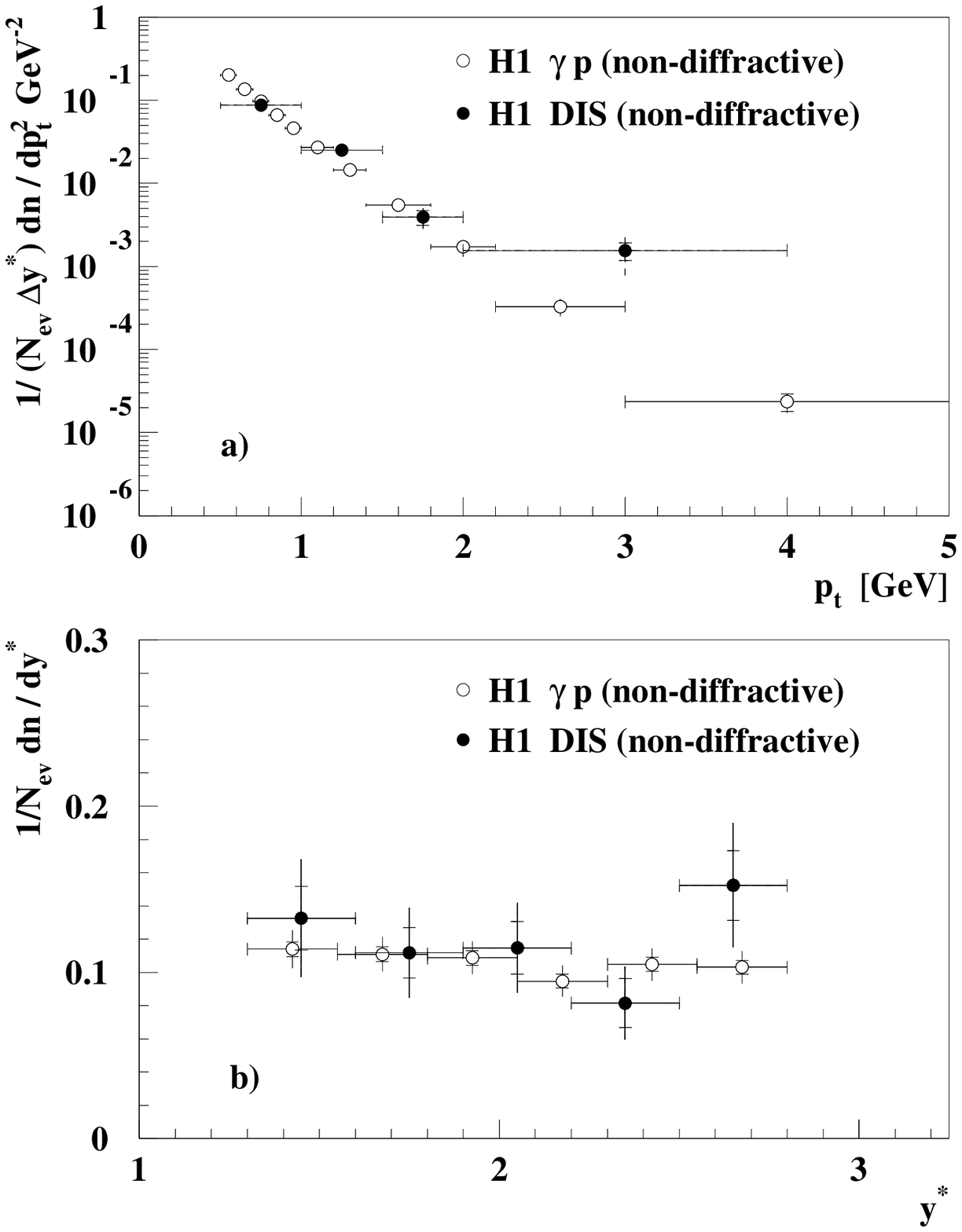,
              height=18cm}
\caption{ $K^0$ rate as a function of $p_t$ (a) and center of mass 
 rapidity $\ystar$ (b) 
 in $\gamma p$ and DIS for non-diffractive events. The data are integrated 
 over $1.3<\ystar<2.8$ in (a), and over $p_t>0.5$~GeV in (b). 
 The inner vertical error bars indicate the statistical error only and the 
 outer error bars the statistical and total systematic error added in 
 quadrature. }
\label{f:pt-ystar-dis-gp-data}
\end{center}
\end{figure}

\begin{figure}[hhhh]
\begin{center}
\epsfig{figure=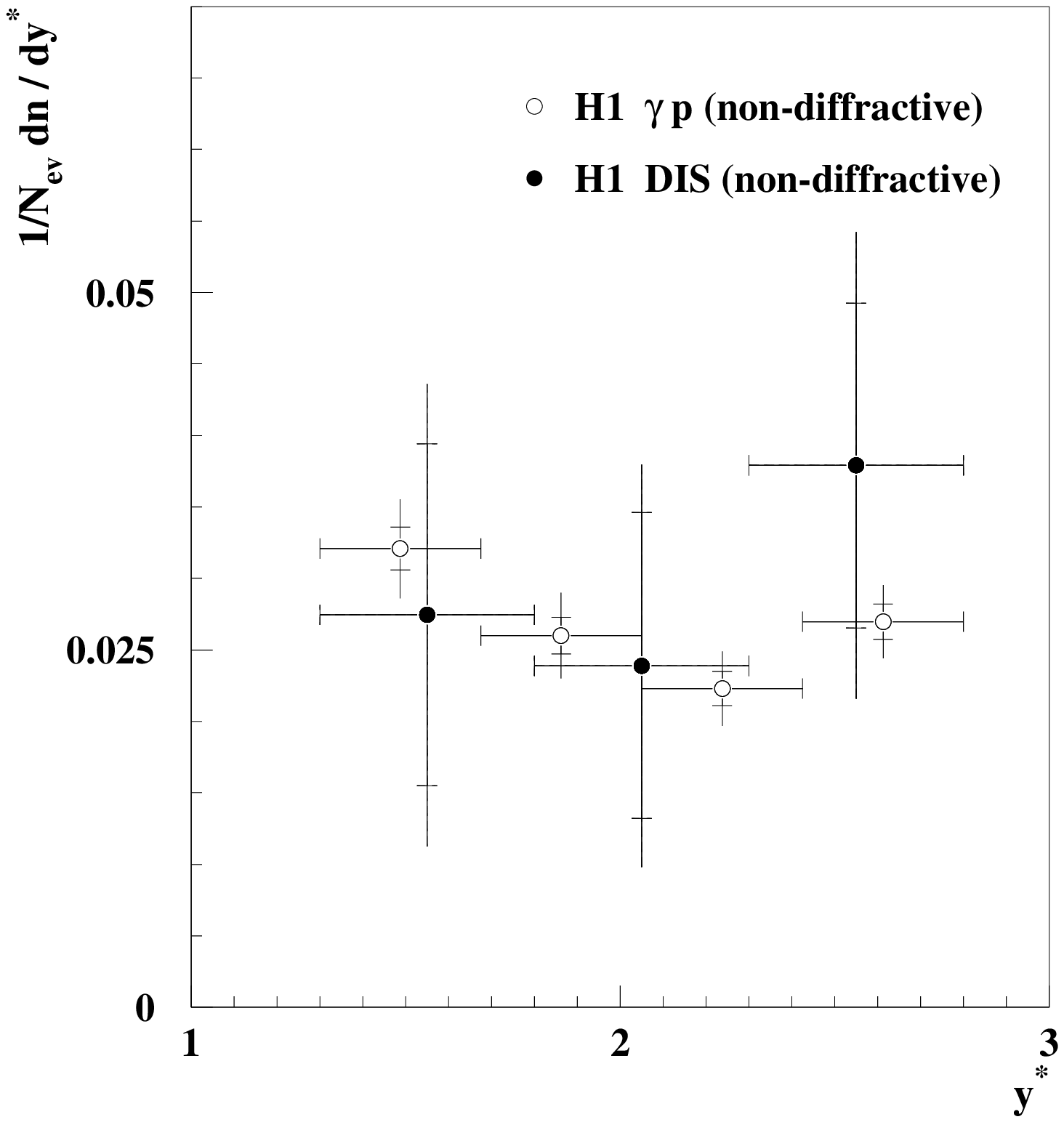,
              height=16cm}
\caption{ $\Lambda$ rate as a function of center of mass 
 rapidity $\ystar$ in $\gamma p$ and DIS for non-diffractive events.
 The data are integrated over $p_t>0.6$~GeV. The inner vertical error bars 
 indicate the statistical error only and the outer error bars the 
 statistical and total systematic error added in quadrature. }
\label{f:ystar-dis-gpl-data}
\end{center}
\end{figure}

A comparison of $K^0$ production rates in $\gamma p$ and DIS is shown in
Fig.~\ref{f:pt-ystar-dis-gp-data} and a similar comparison of $\Lambda$ 
production rates in Fig.~\ref{f:ystar-dis-gpl-data}
for the non-diffractive event samples
defined in section~3.2. The $\gamma p$ and DIS rates
are compatible in $\ystar$ and at low $p_t$.
At high $p_t$, where possible 
differences in the underlying hard subprocesses may 
manifest themselves 
(charm production via photon-gluon fusion, 
hard gluon radiation in DIS, resolved photon
contribution in $\gamma p$), 
the DIS data lie somewhat above the $\gamma p$ data.

\begin{figure}[hhhh]
\begin{center}
\epsfig{file=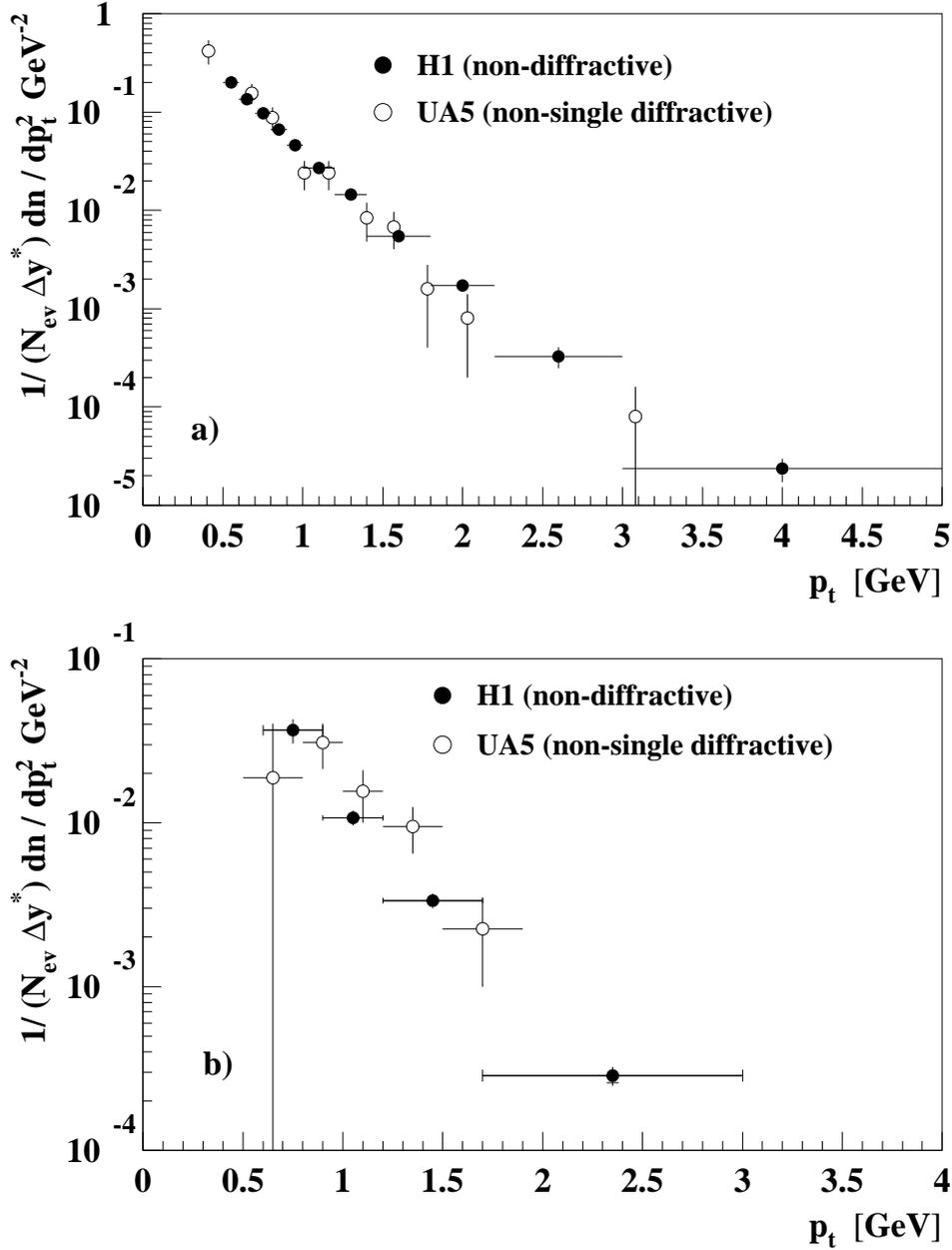,
              height=18cm}
\caption{ $K^0$ rate (a) and $\Lambda$ rate (b) as a function of $p_t$ in
 $\gamma p$ and $p\bar p$ (UA5) for non-diffractive events.
 The H1 data are integrated over $1.3 < \ystar < 2.8$ and the UA5 data are 
 normalized to the same rapidity range. The inner vertical error bars 
 indicate the statistical error only and the outer error bars the statistical 
 and total systematic error added in quadrature. } 
\label{f:pt-ystar-comp-data}
\end{center}
\end{figure}

\begin{figure}[hhhh]
\begin{center}
\epsfig{figure=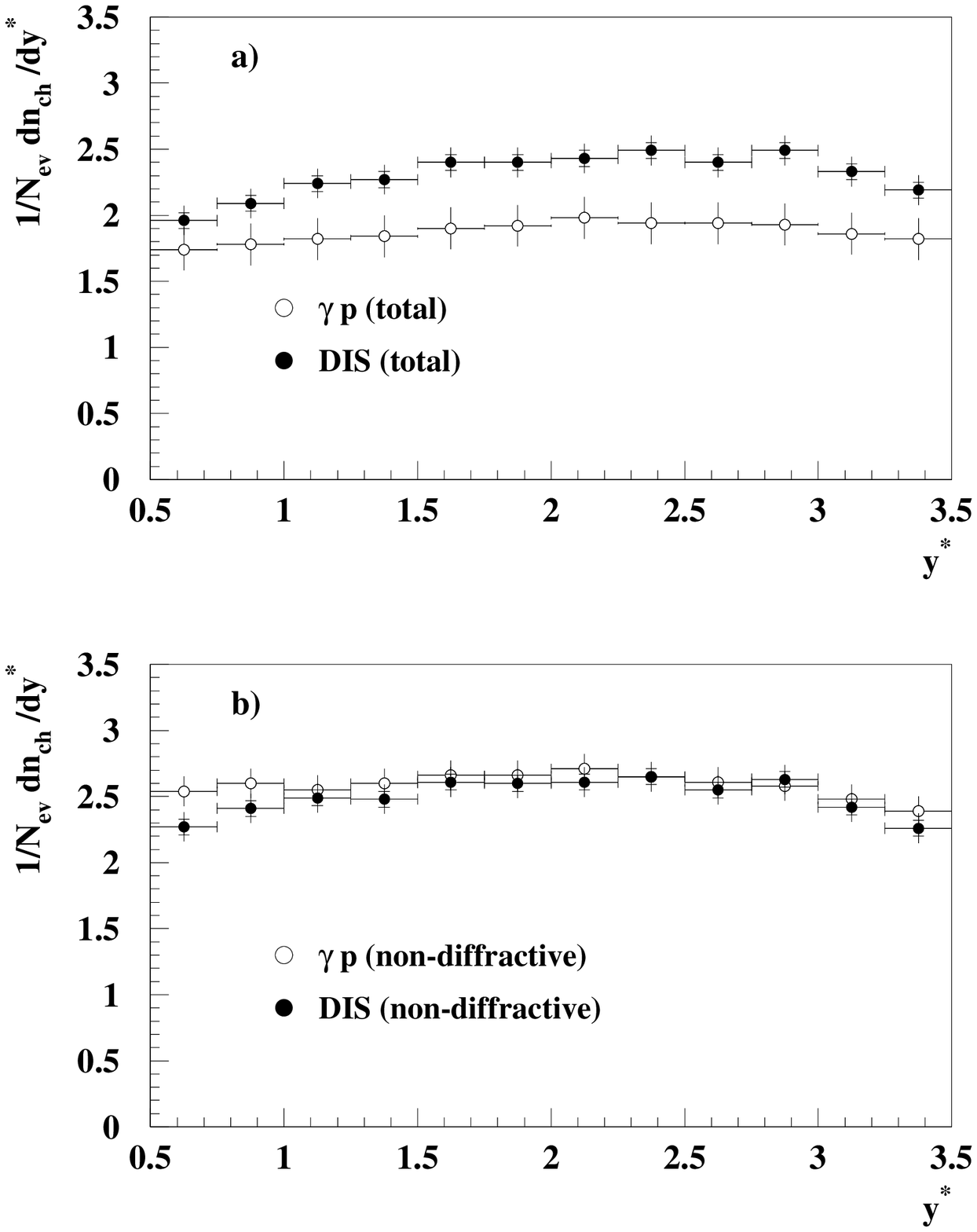,
              height=18cm}
\caption{ The charged-particle center of mass rapidity 
 distribution in DIS and $\gamma p$ for all events (a) and non-diffractive
 events (b). The data have been extrapolated down to $p_t=0$.
 The inner vertical error bars indicate the statistical error only and the 
 outer error bars the statistical and total systematic error added in 
 quadrature. }

\label{f:areta}
\end{center}
\end{figure}
        
It is of interest to investigate if such an agreement in rates extends to
other processes, such as, for instance, $p\bar p$ collisions. 
Figure~\ref{f:pt-ystar-comp-data} shows a comparison of $K^0$  and 
$\Lambda$ production
in $p\bar{p}$ collisions at 200~GeV, obtained by the UA5
collaboration
and the present photoproduction measurement. The UA5 \kzeros\ 
data~\cite{UA52} have been multiplied by  2  to convert the
\kzeros\ rates to $K^0$ and $\bar{K^0}$ rates, and divided by
the width of the rapidity range in which the measurement has been performed
($|\ystar |<2.5$). In the case of the UA5 $\Lambda$ data~\cite{ua5_lambda} the
rapidity range is restricted to $|\ystar |<2$.  
The trigger of the UA5 experiment rejected events in which either only the $p$
or $\bar{p}$ dissociate, as well as elastic events.
This sample, termed non-single diffractive, may be compared to the 
non-diffractive samples of the present measurement as
low-mass diffractive events are removed in both cases. 
It can be seen that the $K^0$ production rates
(Fig.~\ref{f:pt-ystar-comp-data}a) are in good agreement.
In contrast, the $\Lambda$ rates in the photon fragmentation region in
photoproduction are lower than in $p\bar p$ interactions.
This may be attributable to the presence of
an initial baryon in $p\bar p$ in this hemisphere from the incoming proton or
antiproton. A difference between photon and target fragmentation region with
respect to baryon production has also been observed in deep inelastic $\mu N$ 
scattering~\cite{E665}.

Given the observed process independence of the low $p_t$ $K^0$ meson
production rates, it may be asked if such a feature holds for other mesons. 
In order to answer this question, the rate of production of charged particles,
which consist mostly of charged pions and kaons, was investigated in 
$\gamma p$ and DIS. 

For this analysis the charged particles are accepted in the extended
rapidity range $0.5<\ystar<3.5$. At the limits of this range
the geometric acceptance decreases to 80\%.

Figure~\ref{f:areta} shows the comparison of charged-particle rates
observed in DIS and $\gamma p$.
Here the data are extrapolated to $p_t=0$ using samples of Monte Carlo
events which include a full simulation of the effects of the H1 detector.
In Fig.~\ref{f:areta}a, where all events contribute to the center of mass
rapidity spectrum, it can be seen that the rate in DIS is significantly 
higher than in $\gamma p$.
When the (low mass) diffractive events are removed (Fig.~\ref{f:areta}b), the 
rates are  comparable in both samples within errors.
 
The simple quark parton model does not predict an explicit $Q^2$ dependence of
the particle production rates but it can be argued that perturbative QCD 
would introduce such a dependence~\cite{pQCD}.
A small $Q^2$ dependence was reported by EMC~\cite{emc} for charged particles,
but this dependence decreases with increasing $W$.

The present measurement does not indicate a significant change of  
non-diffractive production rates for $K^0$, $\Lambda$
and charged particles in the photon fragmentation region when 
$Q^2$ changes from $Q^2\simeq 0$ to $Q^2\simeq 20$~GeV$^2$. This extends
the observations made at higher $Q^2$ by H1 for charged particles~\cite{Eddie}
and by ZEUS for strange
particles~\cite{zeus-DISk0} that there is little or no dependence of
the average particle multiplicities on $Q^2$. However,
it should be kept in mind that the spectra investigated here are 
dominated by low $p_t$ particles. At higher $p_t$, as mentioned earlier, the
DIS and $\gamma p$ kaon rates show a small discrepancy. This observation is
consistent with results published by H1 showing that for
$p_t>1$~GeV the charged-particle rates increase with increasing 
$Q^2$~\cite{Kuhlen}.

%%%%%%%%%%%%%%%%%%%%%%%%%%%%%%%%%%%
\section{Conclusion}

The $K^0$  and $\Lambda$ production cross sections have been measured
in photoproduction and DIS with the H1 detector at HERA, 
at an average $\gamma p$ center of mass energy of about 
200~GeV, as a function of the transverse momentum and the rapidity, 
in a kinematic range corresponding to the photon fragmentation region. 
No significant enhancement of the non-diffractive $K^0$ and 
$\Lambda$ production rates
is observed when going from $Q^2\simeq 0$ to $Q^2\simeq 20$~GeV$^2$. 
This feature has been found to hold also for the charged-particle 
non-diffractive production rates. At higher $p_t$, however, the $K^0$ rate in
DIS is somewhat higher than in $\gamma p$.
The $K^0$ rate has been found to be similar to that measured in
$p\bar p$ interactions at the same center of mass energy.
However the measured $\Lambda$ rate is lower than in $p \bar p$.
The measured $K^0$ spectra have been found to be broadly consistent
with {\sc Pythia} and NLO predictions, whereas the measured $\Lambda$ spectra 
are significantly underestimated by {\sc Pythia}.

\clearpage
\vskip2ex 
\hbox{\bf Acknowledgments\hss}
We are grateful to the HERA machine group whose outstanding effort made this
experiment possible. We appreciate the immense effort of the engineers and 
technicians who built and maintained the detector. We thank the funding 
agencies for financial support. We acknowledge the support of the DESY 
technical staff. We wish to thank the DESY directorate for the hospitality 
extended to the non-DESY members of the collaboration. We also want to thank 
J.~Binnewies, B.A.~Kniehl and G.~Kramer for fruitful discussions and making 
the results of their calculations available to us.
\vskip0.5ex plus .2pt


\noindent

\begin{thebibliography}{1}
% 
%---------------I N T R O D U C T I O N -----------------------------
%====================================================================

\bibitem{bauer} {\noindent
 T.H.~Bauer, R.D.~Spital, D.R.~Yennie, F.M.~Pipkin,
{\em Rev. Mod. Phys.} {\bf 50}, (1978) {261}; \\
ERRATUM {\it ibid.} {\bf 51}, (1979) {407}. \hfill}


\bibitem{pythia} {\noindent
       T.~Sj\"ostrand, 
       \cpc{82}{1994}{74} \hfill}

\bibitem{Eddie} {\noindent
       H1 Coll., S.~A\"\i d et al.,          
       \zp{C72}{1996}{573} \hfill} 

\bibitem{Kuhlen} {\noindent
       H1 Coll., S.~A\"\i d et al.,          
       \np{B485}{1997}{3} \hfill}

\bibitem{bjkog} {\noindent
	J.D.~Bjorken, J.~Kogut,
        \prev{D8}{1973}{1341} \hfill}

\bibitem{NLO} {\noindent
        J.~Binnewies, B.A.~Kniehl, G.~Kramer,
        {\em Phys. Rev.} {\bf D53}, (1996) {3573}.  \hfill}

\bibitem{schuler} {\noindent
        G.A.~Schuler, T.~Sj\"ostrand,
        \np{B407}{1993}{539} \hfill}

\bibitem{GRV-pion} {\noindent
        W.~Gl\"uck. E.~Reya, A.~Vogt, 
        \zp{C53}{1991}{651} \hfill}

\bibitem{mia} {\noindent
        H1 Coll., S.~A\"\i d et al.,
        \zp{C70}{1996}{17}  \hfill}

\bibitem{E665} {\noindent
        E665 Coll., M.R.~Adams et al.,
        \zp{C61}{1994}{539}  \hfill}

\bibitem{DELPHI} {\noindent
       DELPHI Coll., P.~Abreu et al.,
       \zp{C65}{1995}{587} \hfill}

\bibitem{h1-DISk0} {\noindent
       H1 Coll., S.~A\"\i d et al.,          
       \np{B480}{1996}{3} \hfill}         


\bibitem{GRV-photon/proton} {\noindent
        W.~Gl\"uck. E.~Reya, A.~Vogt, 
        \prev{D46}{1992}{1973} \hfill}


%\bibitem{bkkprivate} {\noindent
%        J.~Binnewies, B.A. Kniehl, G. Kramer,
%         Private communication.   \hfill}

\bibitem{h1det} {\noindent
       H1 Coll., I.~Abt et al., 
 {\it Nucl.\ Instr.\ and Meth.} {\bf A386} (1997) 310; \\
%       H1 Coll., I.~Abt et al., 
% {\it  Nucl.\ Instr.\ and Meth.} {\bf A386} (1997) 348. \hfill}
 {\it  ibid.} {\bf A386} (1997) 348. \hfill}

\bibitem{h1lumi} {\noindent
       H1 Coll., I.~Abt et al., 
       \zp{C66}{1995}{529} \hfill}


\bibitem{t83} {\noindent
       J.~Riedlberger, The H1 Trigger with emphasis on Tracking 
       Triggers, contr. paper to 5th International Conference on Advanced
       Technology and Physics, Como 1994. \hfill}


\bibitem{h1-diffr-gp} {\noindent
       H1 Coll., T.~Ahmed et al.,        
       \np{B435}{1995}{3}  \hfill}


\bibitem{h1-diffr-DIS} {\noindent
       H1 Coll., T.~Ahmed et al., 
       \np{B429}{1994}{477} \hfill}         

\bibitem{karendiss} {\noindent
       K.~Johannsen, Ph.D. thesis, Universit\"at Hamburg,
 DESY FH1-96-01, (1996), unpublished. \hfill}


\bibitem{PDG} {\noindent
        R.M.~Barnett et al., Particle Data Group, {\bf D54} (1996) 412.
	 \hfill}

\bibitem{totalgp} {\noindent
       H1 Coll., S.~A\"\i d et al.,        
       \zp{C69}{1995}{27}  \hfill}


\bibitem{erdmann} {\noindent
	W.~Erdmann, Ph.D. thesis, ETH Z\"urich, No. 11441, (1996),
	unpublished. \hfill}

\bibitem{lumi} {\noindent
        S.~A\"\i d et al., 
       \np{B470}{1996}{3}  \hfill}

\bibitem{flux} {\noindent
        C.F.~Weizs\"acker, 
	{\it Z. Phys.} {\bf 88} (1934) 612; \\
	E.J.~Williams,
        {\em Phys. Rev.} {\bf 45} (1934) 729; \\
	S.~Frixione et al., 
	{\em Phys. Lett.} {\bf B319} (1993) {339}.
  \hfill}

\bibitem{ua5_lambda} {\noindent
        UA5 Coll., R.E.~Ansorge et al., 
       \np{B328}{1989}{36}  \hfill}

\bibitem{UA52} {\noindent
        UA5 Coll., R.E.~Ansorge et al., 
        \zp{C41}{1988}{179} \hfill}

%BASICS OF PERTURBATIVE QCD.

\bibitem{pQCD} {\noindent
Yu.L.~Dokshitzer, V.A.~Khoze, A.H.~Mueller, S.I.~Troyan,
Basics of perturbative QCD,
{\em Ed. Frontieres} (1991). \hfill}

\bibitem{emc} {\noindent
       EMC Coll., M.~Arneodo et al.,          
       {\em Z. Phys.} {\bf C31} (1986) {1}. \hfill}
   
\bibitem{zeus-DISk0} {\noindent
       ZEUS Coll., M.~Derrick et al.,          
       {\em Z. Phys.} {\bf C68} (1995) {29}.  \hfill} 



%%%%%%%%%%%%%%%%%%%%%%%%%%
% --------------------------- Comparison of DIS with gp at HERA
% \bibitem{h1-comp} {\noindent
%       H1 Coll., S.\ A\"\i d et al.,        
%       \pl{B358}{1995}{412} \hfill}

%\bibitem{andrej-paper} {\noindent
%       A.\ Rostovtsev, "contribution to Warsaw" \hfill} 

%\bibitem{lund-string} {\noindent
%       B.\ Andersson et al.,        
%      \prep{97} {1995} {31}  \hfill}  

%\bibitem{aleph} ALEPH Coll., 
%       {\em Zeit. Phys.} {\bf C64}, (1994) {361} . 

%\bibitem{DELPHI} DELPHI Coll., 
%       {\em Zeit. Phys.} {\bf C65}, (1995) {587} . 
   
%\bibitem{delphi_baryon} DELPHI Coll., 
%       {\em Zeit. Phys.} {\bf C67}, (1995) {543} . 


%\bibitem{zeus-diffr-DIS} {\noindent
%       ZEUS Coll., M.\ Derrick et al., 
%       \zp{C68}{1995}{569}
% \hfill} 
 
% POLARIZATION
% --------------------------------------------  PROTON-NUCLEON
%\bibitem{pol:proton-nucleon} {\noindent
% \hfill}
% --------------------------------------------  GP-INTERACTIONS AT 20GEV
%\bibitem{pol:gp-SLAC} {\noindent
%        K.\  Abe  et al. 
%        {\em Phys. Rev.} {\bf D29}, (1984) {1877}  \hfill}

% --------------------------------------------  OMEGA-EXPERIMENT
%\bibitem{pol:gp-OMEGA} {\noindent
%        Omega Experiment \hfill}
%

%--
%
%==========================================================================

%\bibitem{dstar-gp} {\noindent
%       H1 Coll., S.\ A\"{\i}d et al., 
%       DESY preprint 96-55 (1996), to be published in 
%       Nucl.\ Instr.\ and Meth. \hfill}    

%---CHECK THIS REFERENCE OUT

%\bibitem{ijray} {\noindent
%        I.\ Abt, H1 note H1-01/95-420 (1995) \hfill}
%

%---CHECK THIS REFERENCE OUT

%
%\bibitem{UA51} {\noindent
%        G.\ J.\ Alner et al., 
%        \np{B258}{1985}{505} \hfill}
%
%\bibitem{h1chargedp} {\noindent
%       H1 Coll., I.\ Abt et al.,        
%       \pl{B328}{1994}{176} \hfill}
%
%\bibitem{zeuschargedp} {\noindent
%       ZEUS Coll., M.\ Derrick  et al.,        
%       \zp{C70} {1996} 1-16 \hfill}
%
%\bibitem{anishek} {\noindent
%        V.\ V.\ Anisovich, V.\ M.\ Shekhter
%        \np{B55}{1973}{455} \hfill}
%
%\bibitem{anikobr} {\noindent
%        V.\ V.\ Anisovich, M.\ N.\ Kobrinsky
%       \pl{B52}{1974}{217} \hfill}
%
%
%-----------------------------------------------------
%  -- so far unused !!!! -----------------------------
%
%  PHOJET
%\bibitem{phojet} {\noindent
%        R.\ Engel, A.\ Rostovtsev, 
%        H1 note H1-01/95-420 (1995) \hfill}
%
%  PARAMETERIZATION OF TOTAL HADRONIC CROSS SECTIONS
%\bibitem{dola} {\noindent
%        A.\ Donnachie, P.\ V.\ Landshoff,
%        \pl{B296}{1992}{227} \hfill}
%
%\bibitem{Alma} {\noindent
%        B.\ V.\ Batyunya et al. 
%        \zp{C25} {1984} 213-223 \hfill}
%
%  GP-INTERACTIONS AT 20GEV
%\bibitem{SLACgammap} {\noindent
%        K.\  Abe     et al. 
%        Phys.\ Rev.\ D29 1984 1877 \hfill}
%
%   OMEGA-EXPERIMENT
%\bibitem{Omegagammap} {\noindent
%                      Omega Experiment \hfill}
%
%
%\bibitem{pip1} {\noindent
%        D.\  Ljung  et al. 
%        Phys.\ Rev.\ D15 1977 3163 \hfill}
%\bibitem{pip2} {\noindent
%        D.\  Ljung  et al. 
%        Phys.\ Rev.\ D16 1977 2098 \hfill}

%\bibitem{LEPTOMAN} {\noindent
%        G.\ Ingelman, LEPTO Version 6.1
%        Proc. of the Workshop on Physics at HERA, Hamburg 1991,
% vol. 3, p. 1366 \hfill}

%\bibitem{ARIADNE} {\noindent
%        L.\ L\"onnblad,
%        \cpc{71}{1992}{15} \hfill}

%\bibitem{h1particles} {\noindent
%       H1 Coll., S.\ A\"\i d et al.,        
%        \zp{C63}{1994}{377} \hfill}

%\bibitem{zeusparticles} {\noindent
%       ZEUS Coll., M.\ Derrick et al., 
%        \zp{C70}{1996}{1} \hfill}













\end{thebibliography}
\end{document}